\documentclass[10pt]{article}
\usepackage{fa2025}
\usepackage{amsmath}
\usepackage{cite}
\usepackage{url}
\usepackage{graphicx}
\usepackage{color}
\usepackage{siunitx}
\usepackage[utf8]{inputenc}
\usepackage{booktabs}
\let\OLDthebibliography\thebibliography
\renewcommand\thebibliography[1]{
  \OLDthebibliography{#1}
  \setlength{\parskip}{0pt}
  \setlength{\itemsep}{0pt plus 0.3ex}
}
\title{Deep, data-driven modeling of room acoustics: literature review and research perspectives}

\multauthor
{Toon van Waterschoot}{
  Department of Electrical Engineering (ESAT-STADIUS), KU Leuven, Leuven, Belgium
\correspondingauthor{tvanwate@esat.kuleuven.be}{Toon van Waterschoot.}
}

\begin{document}

\maketitle
\begin{abstract}
Our everyday auditory experience is shaped by the acoustics of the indoor environments in which we live. Room acoustics modeling is aimed at establishing mathematical representations of acoustic wave propagation in such environments. These representations are relevant to a variety of problems ranging from echo-aided auditory indoor navigation to restoring speech understanding in cocktail party scenarios. Many disciplines in science and engineering have recently witnessed a paradigm shift powered by deep learning (DL), and room acoustics research is no exception. The majority of deep, data-driven room acoustics models are inspired by DL-based speech and image processing, and hence lack the intrinsic space-time structure of acoustic wave propagation. More recently, DL-based models for room acoustics that include either geometric or wave-based information have delivered promising results, primarily for the problem of sound field reconstruction. In this review paper, we will provide an extensive and structured literature review on deep, data-driven modeling in room acoustics. Moreover, we position these models in a framework that allows for a conceptual comparison with traditional physical and data-driven models. Finally, we identify strengths and shortcomings of deep, data-driven room acoustics models and outline the main challenges for further research.
\end{abstract}

\section{Introduction}\label{sec:introduction}

People spend about 90 \% of their time indoors \cite{klepeis01}, hence our auditory system has been trained to perceive and process sound only after it has been ``shaped'' by the acoustics of our inside living environments. Whereas room acoustics is potentially beneficial for perceptual tasks and experiences including human indoor navigation by means of echolocation \cite{kolarik14,pelegringarcia18} and audience engagement in concert halls \cite{lokki11}, it may also hamper speech understanding \cite{houtgast73} and contribute to the cocktail party effect \cite{bronkhorst00}, in particular for hard-of-hearing people \cite{kollmeier16}.
In addition to this distinction of room acoustics being desired or undesired, problems involving room acoustics can further be classified into forward and inverse problems. Forward problems are aimed at predicting the sound field (or sound signal at one position) in a room, when information about the sound sources (\textit{i.e.}, location, directivity, pressure signal) and the room (\textit{i.e.}, geometry, boundary properties) are given. Inverse problems instead focus on the retrieval of source or room parameters from sound field measurements. 

%
Various perspectives on room acoustics have emerged over the past century. The physical perspective considers room acoustics as a space-time or space-frequency physical phenomenon that can be modeled as an interior boundary value problem, and has been applied primarily to forward 
problems \cite{bilbao16,hamilton17}. The architectural perspective, propelled by the seminal work of Sabine \cite{sabine22}, features compact, empirical or statistical descriptors of room acoustics such as reverberation time (T60), early decay time (EDT), and clarity index (C50) \cite{bradley11}. These descriptors capture the temporal, spectral and/or spatial acoustic behavior of the room at a macroscopic level. Their estimation from sound field measurements is at the core of many inverse 
problems \cite{kendrick08,eaton16}. In the signals and systems perspective, point-to-point acoustic responses within a room are modeled as linear time-invariant (LTI) systems, which can be represented as linear filters \cite{mourjopoulos91,valimaki12} or state-space models \cite{macwilliam24}. This perspective is used for both forward and inverse problems 
considered above, and often involves the processing of acoustic data acquired from microphone signal measurements. Finally, since 2015 a data science perspective on room acoustics has materialized. In this perspective, forward and inverse problems are tackled as predictive learning problems from simulated and/or measured data. As opposed to LTI system modeling, deep learning models involving nonlinear operations are commonly used \cite{michalopoulou21,cobos22,yu24,gotz24,karakonstantis24a}. 


\section{Literature review}\label{sec:review}
\subsection{Concise overview of traditional room acoustics models}
Traditional room acoustics models (\textit{i.e.}, those developed before the deep learning era) can be categorized along two dimensions: the first dimension represents the degree to which models are based on room measurements (“data-driven models”) or on physics first principles (“physical models”), while the second dimension indicates whether the model assumes a geometric or wave-based sound behavior. The leftmost part of Figure \ref{fig:literature} illustrates this categorization and guides the reader through the below concise overview of traditional models.

The architectural and signals and systems perspectives on room acoustics mostly rely on impulse response models, \textit{i.e.}, the room impulse response (RIR) \cite{kuttruff09} and spatial RIR (SRIR) \cite{merimaa05}. Other data-driven models are modal response models, \textit{e.g.}, common-acoustical-pole and zero (CAPZ) \cite{haneda94}, orthonormal basis function (OBF) \cite{vairetti17}, and parallel filter (PF) models \cite{ramo14}. Data-driven models are more effective for room acoustics applications when a physical prior is included in the model structure \cite{vanwaterschoot08}. A geometric prior is used in RIR decomposition models such as the spatial decomposition method (SDM) \cite{tervo13}, whereas wave-based priors are used in wave decomposition models, \textit{e.g.}, wave field analysis (WFA) \cite{berkhout97}, plane-wave decomposition (PWD) \cite{pinto10}, and spherical-harmonic decomposition (SHD) \cite{jarrett17}. Data-driven models incorporating a physical prior based on the boundary integral equation (BIE) \cite{marburg08} include the equivalent source model (ESM) \cite{lee17,antonello17,antonello19} and the boundary integral operator state-space (BIOSS) model \cite{ali25}. Even though the BIE is a wave-based prior, it asymptotically admits a geometric solution \cite{ali23}, hence these models are capable of representing both wave-based and geometric sound behavior. Purely physical models have mainly been used in virtual acoustics and include reflection path models \cite{savioja15} (\textit{e.g.}, image source models (ISM) \cite{allen79}, ray tracing (RT) \cite{krokstad68}, and beam tracing (BT) \cite{funkhouser04}), delay networks (\textit{e.g.}, feedback delay networks (FDN) \cite{jot91}, digital waveguide networks (DWN) \cite{rocchesso97}, and scattering delay networks (SDN) \cite{desena15}), and discretized partial differential equation (PDE) models (\textit{e.g.}, boundary element (BEM) \cite{bai92}, finite element (FEM) \cite{shuku73}, finite difference (FDM) \cite{hamilton17,botteldooren95}, and finite volume models (FVM) \cite{bilbao16}).


\subsection{Purely data-driven deep learning models}
\begin{figure*}
 \centering
 \includegraphics[width=0.91\textwidth]{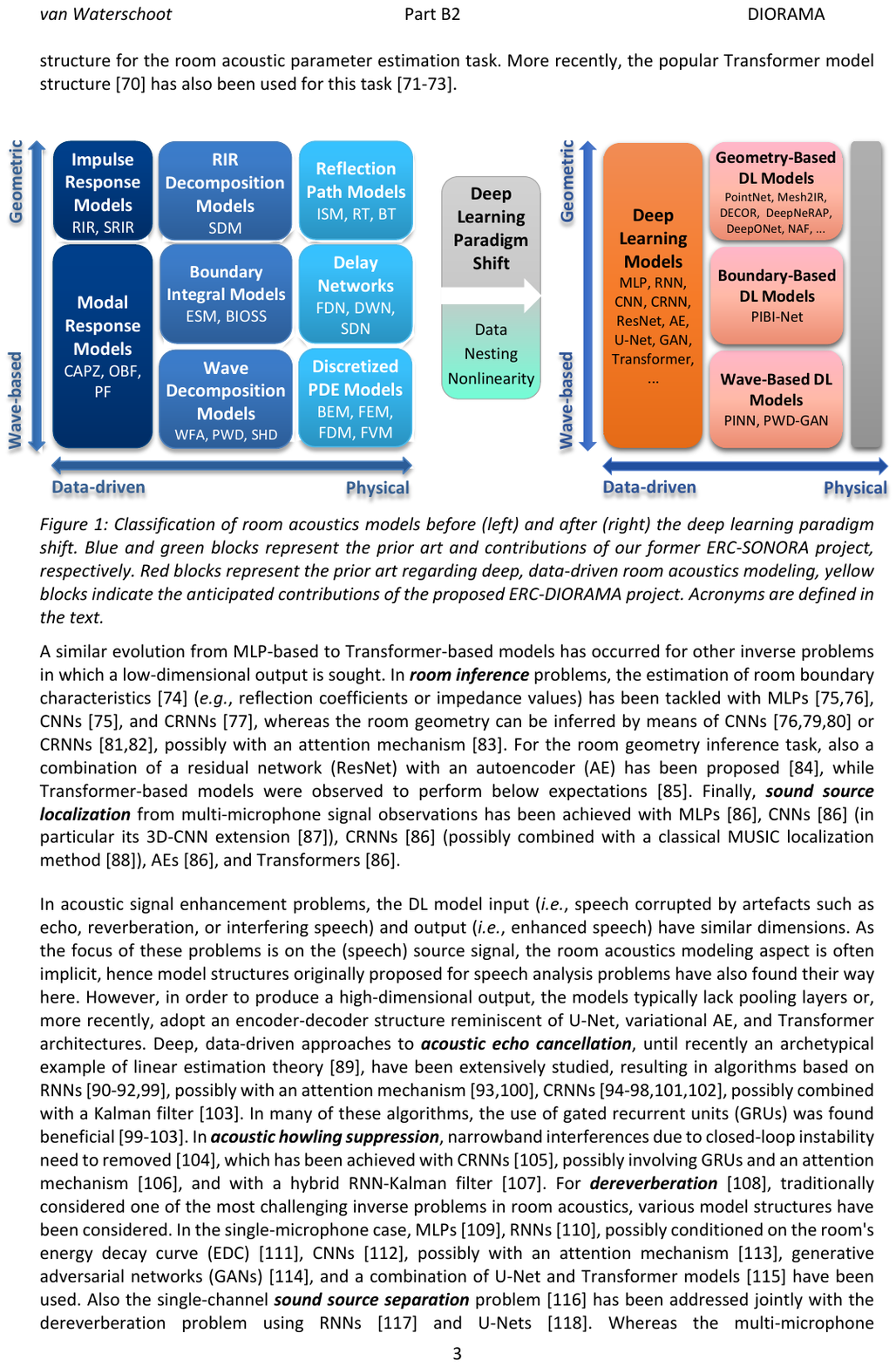}
 \caption{Classification of room acoustics models before (left) and after (right) the deep learning paradigm shift. 
 Acronyms are defined in the text.}
 \label{fig:literature}
\end{figure*}

Over the past decade, many scientific disciplines have witnessed a paradigm shift driven by deep learning (DL). This paradigm shift has also occurred in room acoustics research, and as a consequence, the state of the art has meanwhile fundamentally changed, as illustrated in the rightmost part of Figure \ref{fig:literature}. Not surprisingly, research on deep, data-driven room acoustics modeling has started by considering inverse problems. In these problems, the input is typically a reverberant sound signal from which information on the underlying room acoustics or sound source is to be inferred. Most of the relevant literature considers speech source signals, hence allowing to use DL model structures that have previously shown their merit in speech analysis problems such as automatic speech recognition \cite{yu15,pesoparada15}, speaker identification \cite{bai21}, or speech emotion recognition \cite{tzirakis18,tang21a}. 

The first research efforts in deep, data-driven room acoustics modeling have been focused on inverse problems in which a high-dimensional input (\textit{e.g.}, a reverberant speech signal captured by one or more microphones) is transformed into a low-dimensional output, considering model structures that are often adopted from other DL application areas such as image processing and computer vision. The most widely studied problem in this context is the estimation of room acoustic parameters from reverberant speech \cite{eaton16}: reverberation time (T60) \cite{xiao15,santos16,tang20,bryan20,kehling24,duangpummet22,lopezballester22,ick23,deng20,gotz22,callens20,sanchezlopez21,yang23,wang24a}, clarity index (C50) \cite{pesoparada16,duangpummet22,lopezballester22,callens20,sanchezlopez21,lavechin23,yang23}, direct-to-reverberant ratio (DRR) \cite{bryan20,kehling24,callens20,sanchezlopez21,yang23}, early decay time (EDT) \cite{duangpummet22}, definition (D50) \cite{duangpummet22}, center time (Ts) \cite{duangpummet22}, speech transmission index (STI) \cite{lopezballester22,sanchezlopez21}, speech intelligibility index (SII) \cite{lopezballester22}, room volume \cite{ick23,wang24a,wang24b}, and source distance \cite{neri24}. In a similar context, the classification of rooms from reverberant speech has been investigated \cite{papayiannis20,papayiannis19a}. Even if initially multi-layer perceptron (MLP) model structures were used \cite{xiao15,kehling24}, it was soon realised that long-term temporal characteristics of room reverberation are highly relevant, motivating the use of recurrent neural networks (RNNs) with long-short term memory (LSTM) layers \cite{santos16} or their bidirectional version (BLSTM) \cite{pesoparada16}. Alternatively, convolutional neural networks (CNNs) have been considered \cite{tang20,bryan20,kehling24,duangpummet22,lopezballester22,ick23}, and in particular their combination with recurrent structures into convolutional recurrent neural networks (CRNNs) \cite{deng20,gotz22,callens20,sanchezlopez21,lavechin23}, possibly including an attention mechanism \cite{neri24,papayiannis20}, has become an established model structure for the room acoustic parameter estimation task. More recently, the popular Transformer model structure \cite{vaswani17} has also been used for this task \cite{yang23,wang24a,wang24b}.

A similar evolution from MLP-based over C(R)NN-based to Transformer-based models has occurred for other inverse problems in which a low-dimensional output is sought. In room inference problems, the estimation of room boundary characteristics \cite{antonello15} (\textit{e.g.}, reflection coefficients or impedance values) has been tackled with MLPs \cite{foy21,yu20}, CNNs \cite{foy21}, and CRNNs \cite{papayiannis19b}, whereas the room geometry can be inferred by means of CNNs \cite{yu20,bologni21,liang23a} or CRNNs \cite{poschadel21,tuna23}, possibly with an attention mechanism \cite{bicer24}. For the room geometry inference task, also a combination of a residual network (ResNet) with an autoencoder (AE) has been proposed \cite{yeon23}, while Transformer-based models were observed to perform below expectations \cite{schindler23}. Finally, sound source localization from multi-microphone signal observations has been achieved with MLPs \cite{grumiaux22}, CNNs \cite{grumiaux22} (in particular its 3D-CNN extension \cite{diazguerra20}), CRNNs \cite{grumiaux22} (possibly combined with a classical MUSIC localization method \cite{li23a}), AEs \cite{grumiaux22}, and Transformers \cite{grumiaux22}.

In acoustic signal enhancement problems, the DL model input (\textit{i.e.}, speech corrupted by artefacts such as echo, reverberation, or interfering speech) and output (\textit{i.e.}, enhanced speech) have similar dimensions. As the focus of these problems is on the (speech) source signal, the room acoustics modeling aspect is often implicit, hence model structures originally proposed for speech analysis problems have also found their way here. However, in order to produce a high-dimensional output, the models typically lack pooling layers or, more recently, adopt an encoder-decoder structure reminiscent of U-Net, variational AE, and Transformer architectures. Deep, data-driven approaches to acoustic echo cancellation, until recently an archetypical example of linear estimation theory \cite{vanwaterschoot07}, have been extensively studied, resulting in algorithms based on RNNs \cite{zhang18,zhang20,westhausen21,fazel19}, possibly with an attention mechanism \cite{zhang23a,fazel20}, CRNNs \cite{zhang22,zhang19,cheng21,zhang21a,zhang21b,zhao22a,li23b}, possibly combined with a Kalman filter \cite{liu24}. In many of these algorithms, the use of gated recurrent units (GRUs) was found beneficial \cite{fazel19,fazel20,zhao22a,li23b,liu24}. In acoustic howling suppression, narrowband interferences due to closed-loop instability need to removed \cite{vanwaterschoot11}, which has been achieved with CRNNs \cite{zheng22}, possibly involving GRUs and an attention mechanism \cite{zhang23b}, and with a hybrid RNN-Kalman filter \cite{zhang24}. For dereverberation \cite{naylor10}, traditionally considered one of the most challenging inverse problems in room acoustics, various model structures have been considered. In the single-microphone case, MLPs \cite{zhao18}, RNNs \cite{mack18}, possibly conditioned on the room's energy decay curve (EDC) \cite{bahrman24}, CNNs \cite{luo24}, possibly with an attention mechanism \cite{zhao20}, generative adversarial networks (GANs) \cite{kothapally22}, and a combination of U-Net and Transformer models \cite{donley22} have been used. Also the single-channel sound source separation problem \cite{vincent18} has been addressed jointly with the dereverberation problem using RNNs \cite{delfarah19} and U-Nets \cite{healy21}. Whereas the multi-microphone dereverberation problem seems somewhat underexplored, with the exception of the T60-conditioned MLP model proposed in \cite{wu17}, multi-channel source separation has received more attention with the development of CNN \cite{zermini20}, CRNN \cite{aroudi21}, and U-Net-based models \cite{lluis23,bohlender24}, often designed to work with a spatially preprocessed input such as interaural level/phase differences \cite{zermini20}, ambisonic signal components \cite{lluis23}, and direction-dependent \cite{lluis23} or location-dependent \cite{bohlender24} features. The state-of-the-art SpatialNet approach achieves joint multi-channel speech separation and enhancement by clustering of acoustic transfer functions in a combined Transformer and CNN model structure \cite{quan24}. For the problem of active noise control \cite{kuo99}, MLPs \cite{im23}, CRNNs \cite{zhang21c,zhang23c}, and their combination \cite{cha23} have been considered. Finally, the problem of acoustic matching (\textit{i.e.}, transforming a reverberant sound signal such that it perceptually matches a different room than the one where it was recorded), is typically addressed by means of generative models such as WaveNets \cite{su20} or GANs \cite{chen22}.

Yet another category of problems are those where the desired DL model output is a RIR, a set of RIRs, or a sound pressure field. These problems are fundamentally different from those discussed above, and often require the use of a generative DL model structure. In blind acoustic system identification \cite{gaubitch07}, the aim is to estimate RIRs from reverberant speech observations, and an encoder-decoder structure is generally preferred: the encoder serves to estimate a latent room acoustics representation from the reverberant signal (using CNNs \cite{steinmetz21,ratnarajah24} or ResNets \cite{liao23}), while the decoder adheres to a GAN structure to generate RIRs from the latent embedding \cite{ratnarajah23,steinmetz21,ratnarajah24,liao23}. In the context of room equalization \cite{cecchi17}, a U-Net model structure has been proposed to generate a spatially averaged room transfer function (RTF) from a set of measured RTFs \cite{tuna22}. In sound field reconstruction \cite{verburg18}, one aims to use a set of sound pressure (or RIR) measurements to predict the sound pressure (or RIR) at receiver positions where no measurements have been made. This problem has been addressed with U-Nets \cite{kristoffersen21,pezzoli22}, conditionally invertible neural networks (CINNs) \cite{karakonstantis24b}, and Transformers \cite{qiu24}. In artificial reverberation synthesis \cite{valimaki12}, RIRs are generated for given source and receiver positions and room specifications. Deep, data-driven approaches to this problem include the estimation of FDN parameters from measured RIRs using CNNs \cite{lyster22} and the generation of binaural RIRs for moving receivers with VAEs \cite{sanaguanomoreno24}. Note that even though artificial reverberation synthesis is traditionally a different problem than SFR, the distinction between both problems becomes somewhat ambiguous in a data-driven setting, as the input to both problems consists of room acoustics measurements. Finally, the problem of upmixing RIRs to SRIRs \cite{avendano04} has been tackled with VAEs \cite{yu24} and GANs \cite{xia23}.

\subsection{Deep learning models with physical priors}
Despite their excellent performance for parameter or signal estimation problems, the models discussed above are not suitable (or not optimal) for problems in which RIRs or space-time sounds pressure fields need to be computed, due to the fact that the intrinsic structure of acoustic wave propagation (\textit{e.g.}, representation of time delays, preservation of space-time relations) is not efficiently represented in the model structure. This has given rise to the development of deep, data-driven models that are partially physics-informed, either via geometric or wave-based priors. 

Geometry-based DL models (not to be confused with ``Geometric DL'' which refers to a particular instance of neural networks \cite{gerken23}) have been developed for artificial reverberation synthesis by (1) explicitly encoding the scene geometry with MLPs (PointNet \cite{tang21b}) or GANs with graph convolution layers (Mesh2IR \cite{ratnarajah22a,kelley24}), (2) conditioning the DL model on the source/receiver positions and room geometry with GANs (Fast-RIR \cite{ratnarajah22b}) or (V)AEs \cite{martinsalinas22,martin23}, (3) conditioning the DL model on early reflections with GRU-based CNNs \cite{kim24} or AEs (DECOR \cite{lin24}), or (4) using geometric information only with MLPs (DeepNeRAP \cite{he24}) or neural operators replacing the PDE (DeepONet \cite{borreljensen24}). The strategy of conditioning the DL model on the receiver position has also been used in the context of sound field reconstruction with U-Net \cite{ronchini24} and dynamic kernel \cite{liang24} model structures. A more prominent geometry-based room acoustics model for sound field reconstruction is the Neural Acoustic Field (NAF) \cite{luo22}, which represents the continuous mapping from source/receiver pairs to RIRs by means of MLPs, conditioned on local geometric information present at the source and receiver locations. Extensions of NAF involve the inclusion of boundary information, \textit{i.e.}, boundary geometry (INRAS \cite{su22}) and material properties (NACF \cite{liang23b}), and joint audio-visual scene generation (Few-ShotRIR \cite{majumder22}, NeRAF \cite{brunetto24}, AV-NeRF \cite{liang23c}, SOAF \cite{gao24}). A similar concept named Novel-View Acoustic Synthesis (NVAS) has also been applied to video-aided sound field reconstruction \cite{chen23} and to the inverse problem of joint source localization and recovery \cite{ahn23}. Finally, geometry-based DL has also resulted in novel sound source localization methods by (1) integrating the ray-space transform into CNNs \cite{comanducci20}, (2) exploiting shift-equivariance \cite{berg22} and rotation-equivariance \cite{diazguerraaparicio23} in CNNs, (3) enforcing geometric proximity in the embedding space \cite{tang22}, and (4) combining pairwise networks conditioned on microphone pair positions (Neural-SRP \cite{grinstein23}).

Wave-based DL models are primarily based on physics-informed neural networks (PINNs). The vanilla PINN consists of an MLP-based deep neural network (DNN) which is trained on a loss function that includes a regularization term imposing the acoustic wave equation, and has been successfully applied to artificial reverberation synthesis \cite{borreljensen22} and sound field reconstruction \cite{niebler22,karakonstantis24c}. PINN variations for sound field reconstruction involve the use of trigonometric activation functions (SIREN \cite{pezzoli23,tsunokini24,olivieri24}), regularization with the Helmholtz equation \cite{ma24}, and a deep kernel method regularized by the wave equation \cite{sundstrom24}. Finally, another wave-based DL model for sound field reconstruction consists in the estimation of PWD coefficients by means of a GAN (PWD-GAN \cite{karakonstantis23,fernandezgrande23}).

\section{Research perspectives}\label{sec:perspectives}
From the above literature study, it is clear that deep learning holds great potential for room acoustics modeling. We end this review paper by formulating three prominent perspectives for future research. 

Firstly, data availability remains the first and foremost requirement in the development of deep, data-driven models. Over the past five years, we have witnessed a strong rise in the availability of high-quality, large-scale, and open-access datasets of room acoustics measurements in diverse conditions and measurement setups, \textit{e.g.}, \cite{kristoffersen21,szoke19,cmejla20,carlo21,koyama21,zhao22b,dietzen23,fejgin23,chen24,chesworth24,yang24,kujawski24,miotello24,fragner24,stolz24,damiano24}. The main challenge in developing additional datasets is to reconcile the conflicting requirements of designing setups that correspond to realistic audio scenes (\textit{e.g.}, with moving and directional sources and microphones) while allowing for accurate data labeling (\textit{e.g.}, in terms of source and microphone positions and orientations).

Secondly, a more fundamental understanding is required of why deep, data-driven models appear to be highly suitable for room acoustics modeling. Two key elements of deep, data-driven models are their nested model structure consisting of composed functions referred to as layers, and the nonlinear activation functions used in these layers. It is not well understood how nesting and nonlinearity contribute to the accurate and efficient modeling of room acoustics, in particular as these properties seem to contradict the widely accepted premise that room acoustics can be modeled as a linear, time-invariant process. 

Thirdly, upon comparing our classification of traditional models with deep, data-driven models for room acoustics in Figure \ref{fig:literature}, there is an apparent gap between geometry-based and wave-based deep, data-driven room acoustics models. Given the asymptotic geometric interpretation of the wave-based BIE discussed above \cite{ali23}, the key to combining geometric and wave-based information may lie in the inclusion of boundary information both in the model structure and in the model training strategy of DL models. A first and promising result in this direction consists in the use of physics-informed boundary integral networks (PIBI-Nets) for sound field reconstruction \cite{damiano25}.

\begin{scriptsize}
\section{Acknowledgements}
This research work was carried out at the ESAT Laboratory of KU Leuven, in the frame of KU Leuven internal funds C14/21/075 and C3/23/056, FWO projects G0A0424N and S005525N, and the AI Research Program of the Flemish Government. The scientific responsibility is assumed by its authors.

\bibliography{fa2025_toon}
\end{scriptsize}

%
%
%

\end{document}